\documentclass[aps,prl,twocolumn,superscriptaddress]{revtex4-1}

\usepackage{mhchem,graphicx,longtable,xfrac}
\usepackage[colorlinks=true,citecolor=blue,linkcolor=blue,breaklinks=true]{hyperref}
\usepackage{amssymb}
\usepackage{graphicx}
\usepackage{amsmath}

\usepackage{times}
\usepackage{color}
\usepackage{bm}

\begin{document}

\newcommand {\beq} {\begin{equation}}
\newcommand {\eeq} {\end{equation}}
\newcommand {\bqa} {\begin{eqnarray}}
\newcommand {\eqa} {\end{eqnarray}}
\newcommand {\ba} {\ensuremath{b^\dagger}}
\newcommand {\Ma} {\ensuremath{M^\dagger}}
\newcommand {\psia} {\ensuremath{\psi^\dagger}}
\newcommand {\psita} {\ensuremath{\tilde{\psi}^\dagger}}
\newcommand{\lp} {\ensuremath{{\lambda '}}}
\newcommand{\A} {\ensuremath{{\bf A}}}
\newcommand{\Q} {\ensuremath{{\bf Q}}}
\newcommand{\kk} {\ensuremath{{\bf k}}}
\newcommand{\qq} {\ensuremath{{\bf q}}}
\newcommand{\kp} {\ensuremath{{\bf k'}}}
\newcommand{\rr} {\ensuremath{{\bf r}}}
\newcommand{\rp} {\ensuremath{{\bf r'}}}
\newcommand {\ep} {\ensuremath{\epsilon}}
\newcommand{\nbr} {\ensuremath{\langle ij \rangle}}
\newcommand {\no} {\nonumber}
\newcommand{\up} {\ensuremath{\uparrow}}
\newcommand{\dn} {\ensuremath{\downarrow}}
\newcommand{\rcol} {\textcolor{red}}



\title{Antiferromagnetism and crystalline-electric field excitations in tetragonal NaCeO$_2$}

\author{Mitchell M. Bordelon}
\affiliation{Materials Department, University of California, Santa Barbara, California 93106, USA}

\author{Joshua D. Bocarsly}
\affiliation{Materials Research Laboratory, University of California, Santa Barbara, California 93106, USA}

\author{Lorenzo Posthuma}
\affiliation{Materials Department, University of California, Santa Barbara, California 93106, USA}

\author{Arnab Banerjee}
\affiliation{Neutron Scattering Division, Oak Ridge National Laboratory, Oak Ridge, TN 37831, USA}

\author{Qiang Zhang}
\affiliation{Neutron Scattering Division, Oak Ridge National Laboratory, Oak Ridge, TN 37831, USA}

\author{Stephen D. Wilson}
\email[]{stephendwilson@ucsb.edu}
\affiliation{Materials Department, University of California, Santa Barbara, California 93106, USA}


\date{\today}

\begin{abstract}
	
	We investigate the crystal structure, magnetic properties, and crystalline-electric field of tetragonal, $I4_1/amd$, NaCeO$_2$.  In this compound, Ce$^{3+}$ ions form a tetragonally elongated diamond lattice coupled by antiferromagnetic interactions ($\Theta_{CW} = -7.69$ K) that magnetically order below $T_N = 3.18$ K. The Ce$^{3+}$ $J = 5/2$ crystalline-electric field-split multiplet is studied via inelastic neutron scattering to parameterize a $J_{eff} = 1/2$ ground state doublet comprised of states possessing mixed $|m_z \rangle$ character. Neutron powder diffraction data reveal the onset of $A$-type antiferromagnetism with $\mu=0.57(2)$ $\mu_B$ moments aligned along the $c$-axis. The magnetic structure is consistent with the expectations of a frustrated Heisenberg $J_1$\---$J_2$ model on the elongated diamond lattice with effective exchange values $J_1 > 4 J_2$ and $J_1 > 0$.

\end{abstract}

\maketitle

\section{I. Introduction}

Due to their highly localized moments and their ability to incorporate within a wide variety of frustrated lattice geometries, studying magnetic interactions in lanthanide-based materials is a rich testbed for numerous model Hamiltonians. Perhaps most prominently, the rare earth pyrochlores of general formula $Ln_2M_2O_7$ ($Ln =$ lanthanide; $M = $ metal or metalloid) display unusual magnetic phases like (quantum) spin ice and (quantum) spin liquids \cite{Lee, balents4, balents5, witczak6, zhou7, cava-broholm, lee-balents, gchen, buessen-trebst, bramwell1998frustration, harris1996frustration, ramirez1999zero, bramwell2001spinice, ross2011quantum, moessner1998properties, canals1998pyrochlore}, order by disorder phenomena \cite{bergman-balents, bernier2008quantum, savary2011impurity, bellier2001frustrated}, and hidden multipolar order \cite{anand2015observation, lhotel2015fluctuations, takatsu2016quadrupole, sibille2020quantum, tomiyasu2012emergence, li2017symmetry, benton2016quantum, petit2016observation, mauws2018dipolar, gaudet2019quantum}. This results from their three dimensional network of frustrated trivalent lanthanide moments. More recently, layered trivalent lanthanide triangular lattice compounds also have shown promise to exhibit model quantum disordered states, for example the materials YbMgGaO$_4$ \cite{zhang_mourigal, zhu_chernyshev, li_chen, li_chen2, li_zhang, li_zhang2, li_zhang3, li_zhang4, maksimov_chernyshev, paddison_mourigal, shen_zhao, xu_li} and NaYb$X$$_2$ ($X = $chalcogenide) \cite{baenitz_doert, paper1NYO, bordelon2020spin, ding_tsirlin, liu_zhangAMX2, ranjith_baenitz, ranjith_baenitz2, sarkar_klauss, sichelschmidt_doert}. 

The recently reported $I4_1/amd$ tetragonal structure of LiYbO$_2$ has been suggested as another example of an antiferromagnetically frustrated lanthanide system in a three dimensionally connected lattice \cite{LYOpaper}. In this material, the $J_{eff} = 1/2$ Yb$^{3+}$ moments decorate a bipartite lattice that can be interpreted as an extreme elongation of the magnetic diamond lattice. In fact, the simplest magnetic model for the diamond lattice, the Heisenberg $J_1$\---$J_2$ model in Equation \ref{eq:J1J2}, was adapted to model the incommensurate spiral magnetic order observed in LiYbO$_2$, where the propagation wave vector is directly ascertained by the ratio of $J_2/|J_1|$ \cite{LYOpaper}.

\begin{equation}
\label{eq:J1J2}
H = J_1\sum_{<i,j>} {\bf{S_i}}\cdot{\bf{S_j}} + J_2\sum_{<<i,j>>} {\bf{S_i}}\cdot{\bf{S_j}}
\end{equation}

\begin{figure}[t]
	\includegraphics[scale=.45]{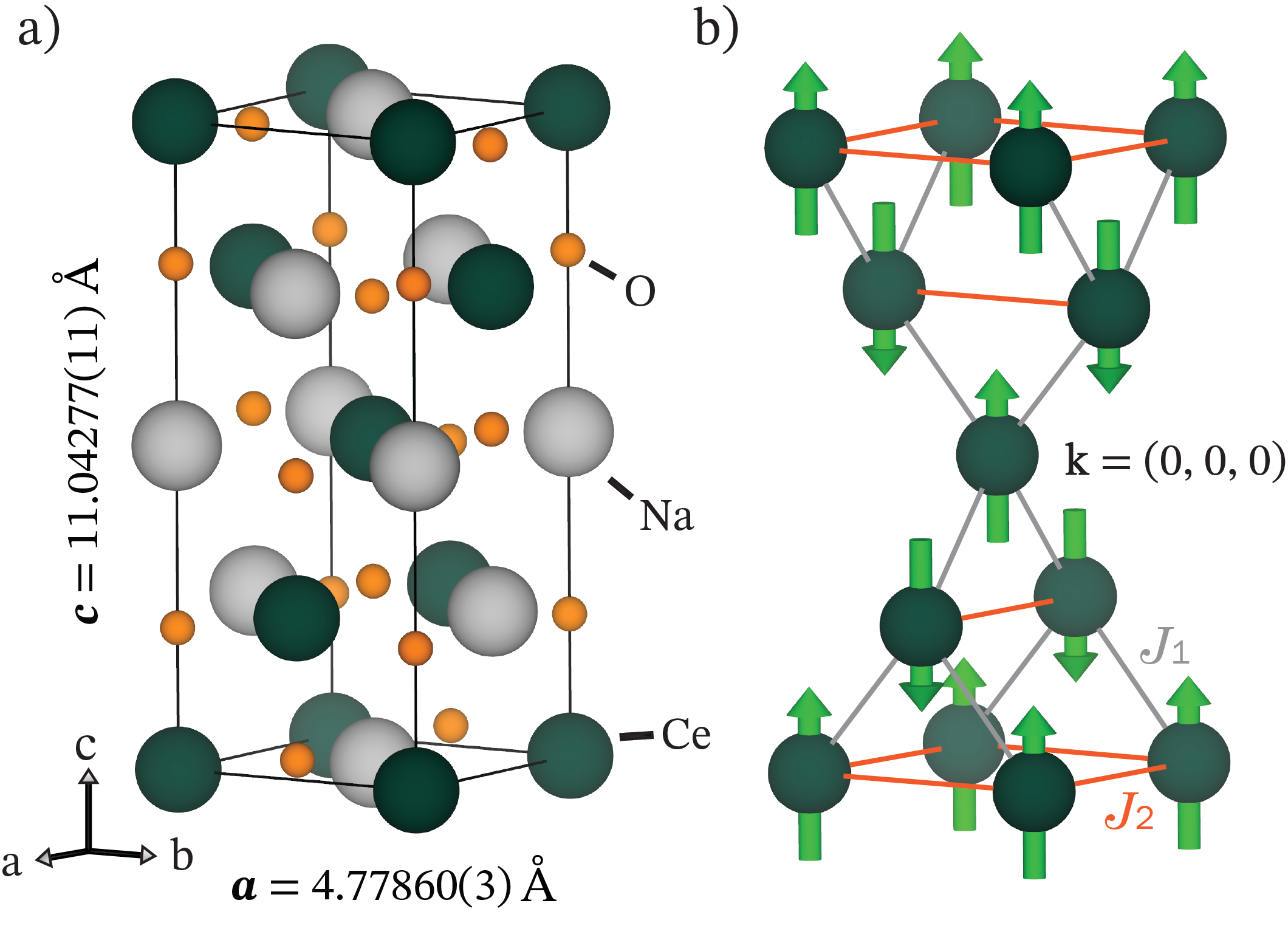}
	\caption{a) Schematic of the 1.5 K refined crystal structure from neutron powder diffraction in origin setting 2 of $I4_1/amd$. Trivalent Ce ions reside in $D_{2d}$ distorted CeO$_6$ octahedra that form a bipartite, tetragonally-elongated diamond lattice. b) The corresponding magnetic structure below $T_N = 3.18$ K contains $0.57(2) \mu_B$ Ce moments oriented parallel to the $c$-axis. Magnetic interactions along $J_1$ and $J_2$ are at distances 3.65105 \AA \ and 4.77860 \AA, respectively, where both contain a superexchange pathway with one oxygen atom.}
	\label{fig:figstruct}
\end{figure}

How the ground state evolves as anisotropies, lattice distances, and exchange pathways are modified in these stretched diamond lattice $I4_1/amd$ materials within the $ALnX_2$ family ($A =$ alkali; $Ln =$ lanthanide; $X =$ chalcogen) remains unexplored. The spin-orbit entangled lanthanide moments can host a variety of local anisotropies, which can result in a diversity of behaviors across the $Ln$ ion materials on the same crystallographic framework. Furthermore, the variation in the $Ln$ ion and $A$-site cation sizes should alter the ratio of $J_2/|J_1|$ and sample different sections of the phase space of the frustrated Heisenberg model. For instance, it is known that when $J_1$ and $J_2$ compete, spiral order emerges as the ground state \cite{LYOpaper}; however in the limit where either interaction dominates, conventional N\'eel or ferromagnetic order should arise.

\begin{figure*}[t]
	\includegraphics[width=\textwidth*10/10]{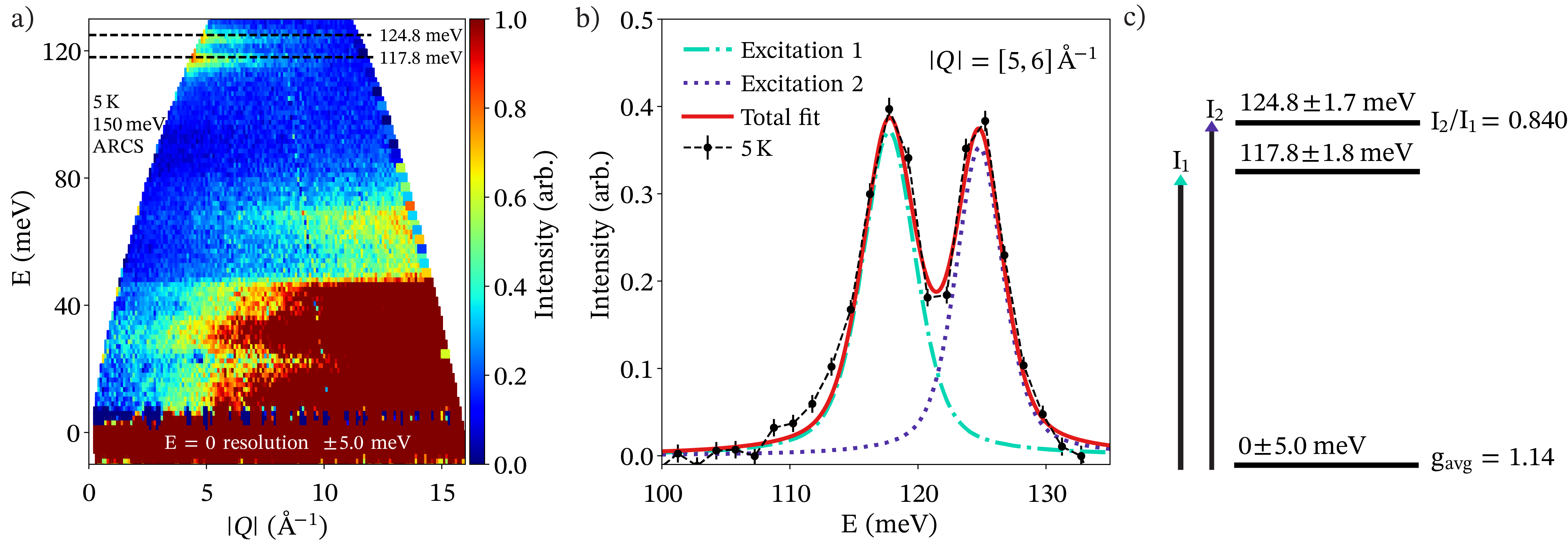}
	\caption{a) Inelastic neutron scattering data $I(\bf{Q},\hbar \omega)$ collected at 5 K and $E_i = 150$ meV on the ARCS spectrometer with full width at half maximum energy resolution at the elastic line of 5.0 meV. Two CEF excitations within the $J = 5/2$ Ce$^{3+}$ ground state manifold are indicated by dashed black lines at 117.8 and 124.8 meV. b) $\bf{Q}$-integrated cut through the data with an empirical linear background subtracted and peak shapes modeled with a pseudo Voigt function. c) Schematic of the energy levels of $J = 5/2$ Ce$^{3+}$ manifold split into three Kramers doublets drawn at 0, 117.8, and 124.8 meV with corresponding full width at half maximum energy resolution of the spectrometer at each energy transfer.}
	\label{fig:figCEF}
\end{figure*}

In this paper, we study the magnetic behavior and ground state of NaCeO$_2$ (Figure \ref{fig:figstruct}a), which shares the same crystallographic structure as LiYbO$_2$ in the $ALnX_2$ materials family. We show that the Ce$^{3+}$ moments are described by a well separated $J_{eff} = 1/2$ Kramers ground state doublet and inelastic neutron scattering data determine the level structure for the split $J=5/2$ ground state multiplet. Low temperature magnetization, heat capacity and neutron diffraction measurements reveal that NaCeO$_2$ magnetically orders below $T_N = 3.18$ K into an A-type antiferromagnetic phase shown in in Figure \ref{fig:figstruct}b.  This ground state matches the predictions of a $J_1-J_2$ Heisenberg model with a dominant nearest-neighbor $J_1 > 0$ interaction. 

\section{II. Methods}

\subsection{Sample preparation}

As has been previously reported, NaCeO$_2$ can be prepared by reducing CeO$_2$ in the presence of liquid sodium \cite{mignanelli1981investigation, barker1981polymorphism, barker1995reaction}. Here, we synthesized polycrystalline samples of NaCeO$_2$ from a 1.1:1.0 molar ratio of Na (99.95\% Alfa Aesar) and CeO$_2$ (99.99\% Alfa Aesar), respectively, in sealed 316 stainless steel tubes, following a similar synthesis method previously reported for NaTiO$_2$ \cite{wu2015natio}. Reagents were inserted into the steel tubes in an argon-filled glove box with oxygen and water content below 0.5 ppm, and the tube was sealed within the glove box. The capped steel tubes were then placed into a tube furnace under held under vacuum at 1000 $^{\circ}$C for three days and were subsequently opened in a glove box. The resulting green NaCeO$_2$ powder rapidly degrades when in contact with air and moisture, and therefore it was handled solely in dry, inert environments. Sample composition was verified via x-ray diffraction measurements under Kapton film on a Panalytical Empyrean powder diffractometer with Cu-K$\alpha$ radiation, and data were analyzed using the Rietveld method in the Fullprof software suite \cite{rodriguezfullprof}.
  
\subsection{Magnetic susceptibility \& heat capacity}

Bulk magnetic properties of NaCeO$_2$ were measured on a Quantum Design Magnetic Properties Measurement System (MPMS3) with a 7 T magnet and a Quantum Design Physical Properties Measurement System (PPMS) with at 14 T magnet and vibrating sample magnetometer. The magnetization of NaCeO$_2$ was measured from 2$-$300 K under zero-field-cooled (ZFC) and field-cooled (FC) conditions with an applied field of $\mu_0$H $= 50$ Oe in the MPMS3. Isothermal magnetization data were collected at 2, 20, 100, and 300 K in the PPMS under fields up to $\mu_0$H $= 14$ T. Specific heat measurements were also measured within the PPMS, and heat capacity data were collected between 2 K and 300 K on pressed NaCeO$_2$ in external fields of $\mu_0$H $= 0, 3, 5,$and $9$ T.

\subsection{Neutron scattering} 

Elastic neutron powder diffraction data were obtained from the POWGEN diffractometer at the Spallation Neutron Source in Oak Ridge National Laboratory. A powder sample in a vanadium canister was loaded into a helium-flow cryostat where data were collected at 1.5 K and 10 K in the instrument's Frame 2 (centered at $\lambda = 1.5$ \AA) and Frame 3 (centered at $\lambda = 2.665$ \AA). Time-of-flight diffraction patterns from both frames were co-refined with the Topas Academic software suite \cite{topas} for both magnetic and structural phases. Magnetic structural refinements were set up with the aid of ISODISTORT \cite{isodistort, isodisplace}. 

Inelastic neutron scattering data were obtained on the wide Angular-Range Chopper Spectrometer (ARCS) at the Spallation Neutron Source in Oak Ridge National Laboratory. Incident neutrons of energy $E_i = 150$ meV (Fermi 2, Fermi frequency 600 Hz) were used at $T = 5$ K. Contributions from the aluminum sample can were removed by measuring an empty canister at the same $E_i$ and $T$. Crystalline-electric field analysis utilized a $Q$-integrated energy cut ($E$-cut) of the $E_i = 150$ meV INS data between $|Q| = [5, 6]$ \AA$^{-1}$. Peaks were fit to a Pseudo-Voigt function that approximates the beam shape of the instrument.

\subsection{Crystalline-electric field analysis} 

Analysis of the crystalline-electric field (CEF) level scheme followed the generic procedure presented in Bordelon et al. \cite{bordelon2020spin} for NaYbO$_2$ and Bordelon et al. \cite{LYOpaper} for LiYbO$_2$. A synopsis of the procedure is detailed below with changes reflected for NaCeO$_2$.  NaCeO$_2$ consists of trivalent Ce$^{3+}$ ions with total angular momentum $J = 5/2$ ($L = 3$, $S = 1/2$) ($4f^1$). In the Ce ion's local $D_{2d}$ setting, the $J = 5/2$ manifold can be maximally split into a set of three Kramers doublets. The splitting can be estimated by a point charge (PC) model of ions surrounding a central Ce ion in the crystal field interface of Mantid Plot \cite{Mantid}. Table \ref{tab:tabCEF} displays three different coordination shell PC models utilizing the refined structural parameters for NaCeO$_2$ from neutron powder diffraction. 

The minimal Hamiltonian with CEF parameters $B_n^m$ and Steven's operators $\hat{O}_m^n$ \cite{StevensOperators} in $D_{2d}$ symmetry for Ce$^{3+}$ is as follows:

\begin{equation}
	\label{eq:CEF}
	H_{CEF} = B_2^0 \hat{O}_2^0 + B_4^0 \hat{O}_4^0 + B_4^4 \hat{O}_4^4
\end{equation}

Diagonalizing the CEF Hamiltonian returns the multiplet's energy eigenvalues, relative probabilities (intensities) of the intramultiplet transitions, wave function basis vectors for each CEF-split doublet, and a powder averaged $g_{avg}$ factor for the ground state doublet. The PC model approximations were used as starting points for fitting the $E$-cut of INS data by following a numerical error minimization procedure \cite{bordelon2020spin}.

\section{III. Experimental Results}

\subsection{Crystalline-electric field excitations}

\begin{table}[h]
	
	\caption{Point charge (PC) models and the CEF fit for NaCeO$_2$ obtained by minimizing observed parameters from $E_i$ = 150 meV inelastic neutron scattering data and the powder-averaged $g_{avg}$ factor from the magnetic structure. Three PC models represent three coordination shells from a central Ce ion of increasing size, where the first includes O$^{2-}$ ions only, the second has O$^{2-}$ and Na$^{+}$ ions, and the third has O$^{2-}$, Na$^{+}$, and nearest neighbor Ce$^{3+}$ ions.}
	\begin{tabular*}{9.0cm}{l|llll|l|llll}
		\hline
		& $E_1$   & $E_2$    & $\frac{I_2}{I_1}$   & $g_{avg}$  & $\chi^2$ & $B_2^0$      &  $B_4^0$   &  $B_4^4$           \\ \hline
		PC (2.5 \AA)  & 86.2 & 122.7 & 0.763 & 1.46 & 8.61 & -2.4869 & 0.2766 & 1.4544      \\
		PC (3.5 \AA)  & 104.6 & 223.5 & 0.243  & 2.07 & 80.73 & -10.2981 & 0.2645 & 1.5953 		\\
		PC (3.7 \AA)  & 53.7 & 179.3  & 0.744  & 1.45 & 58.78 & 7.4129 & 0.2943 & 1.5249        \\
		
		Fit  & 117.9 & 124.8 & 0.844 & 1.15 & 0.0003  & 0.9254 & 0.3701 & 1.3928      \\ \hline 
		Observed   & 117.8 & 124.8  & 0.840 & 1.14  & & &  &  &   
	\end{tabular*}
	
	
	\hskip+0.1cm
	\begin{tabular*}{9.025cm}{ll}
		\hline
		Fit wave functions:     &  \\
		$| \omega_{0, \pm}\rangle =$  $0.949|\pm3/2\rangle -0.316|\mp5/2\rangle$ &\\		
		$| \omega_{1, \pm}\rangle =$  $1|\pm1/2\rangle$ &\\
		$| \omega_{3, \pm}\rangle =$  $0.316|\mp3/2\rangle +0.949|\pm5/2\rangle$ &\\ \hline
	\end{tabular*}
	\label{tab:tabCEF}
	
\end{table}

Inelastic neutron scattering data for NaCeO$_2$ are displayed in Figure \ref{fig:figCEF} and were used to determine the $J = 5/2$ CEF scheme of trivalent Ce ions. In Figure \ref{fig:figCEF}a, two resolution-limited excitations out of ground state doublet appear and are centered at $E_1$=117.8 meV and $E_2=$ 124.8 meV. An energy cut through $I(\bf{Q},\hbar \omega)$ plotted in Figure \ref{fig:figCEF}b shows that these excitations are sharp with lifetimes limited by the instrumental resolution at $E_i = 150$ meV. The CEF strongly separates the ground state Kramers doublet in NaCeO$_2$ from other states, isolating it as a $J_{eff} = 1/2$ state at low temperatures. The entire $J = 5/2$ CEF manifold was fit with extracted parameters from the data in Figure \ref{fig:figCEF}, and the results are shown in Table \ref{tab:tabCEF} and represented schematically in Figure \ref{fig:figCEF}c. Initial guesses for the fitting procedure were based off of point charge (PC) calculations in Table \ref{tab:tabCEF}. A powder averaged $g$-factor $g_{avg} = \sqrt (1/3( g_{//}^2 + 2g_{\perp}^2))$ was utilized as another constraint on the model and was extracted from the ordered moment obtained from the magnetic structure detailed later in this paper (0.57 $\mu_B = g_{avg} \ J_{eff} \ \mu_B$). The final fit of the CEF-scheme revealed a ground state wave function with mixed $m_z=3/2$ and $m_z=5/2$ character and a moderately anisotropic $g$ tensor ($g_{//} = 1.41$ and $g_{\perp} = 1.00$).

\subsection{Heat capacity, magnetization, and susceptibility results}

\begin{figure*}[!ht]
	\includegraphics[width=\textwidth*10/10]{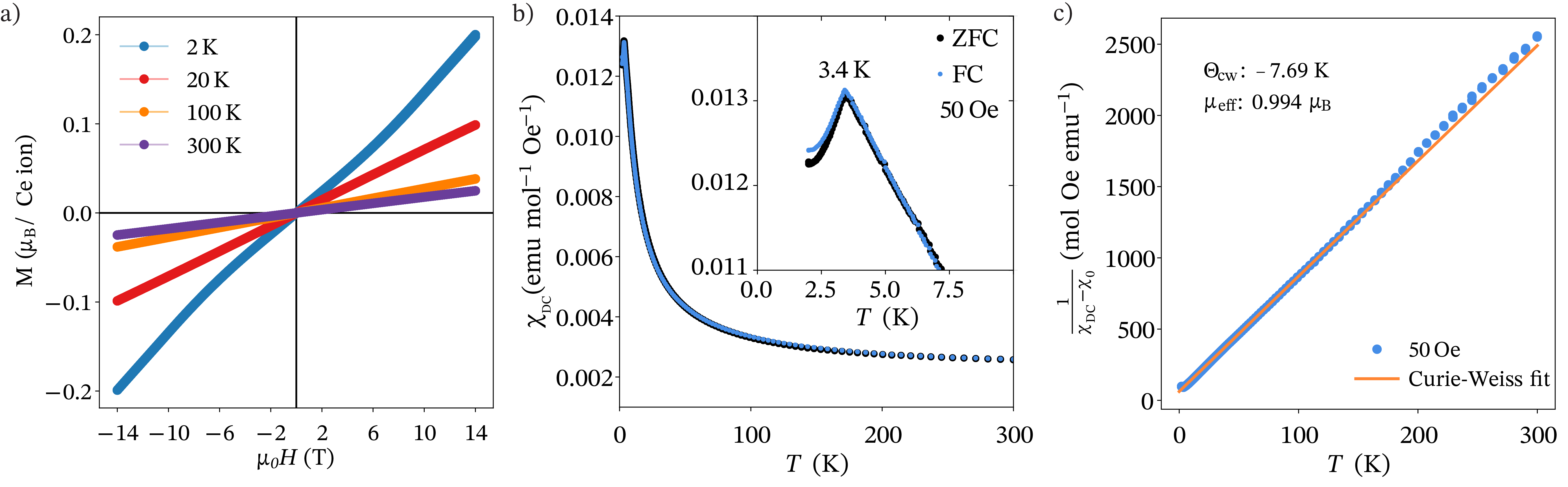}
	\caption{a) Isothermal magnetization of NaCeO$_2$ collected at $T = 2, 20, 100, 300$ K. Below 2 K, NaCeO$_2$ is magnetically ordered and reaches $0.2 \mu_B$/Ce at $\mu_0H = 14$ T. b) Magnetic susceptibility collected under zero field cooled (ZFC) and field cooled (FC) conditions with $\mu_0H = 50$ Oe. Inset: Zoom in of the magnetic susceptibility at low temperature showing a splitting of ZFC and FC below 3.4 K. c) Inverse susceptibility of NaCeO$_2$ fit between $50<T<200$ K to the Curie-Weiss relationship where $\mu_{eff} = g_{avg} \sqrt{J(J+1)}$, corresponding to $g_{avg} =$ 1.15 with $J = 1/2$.}
	\label{fig:figmagnet}
\end{figure*}

\begin{figure}[]
	\includegraphics[scale=.5]{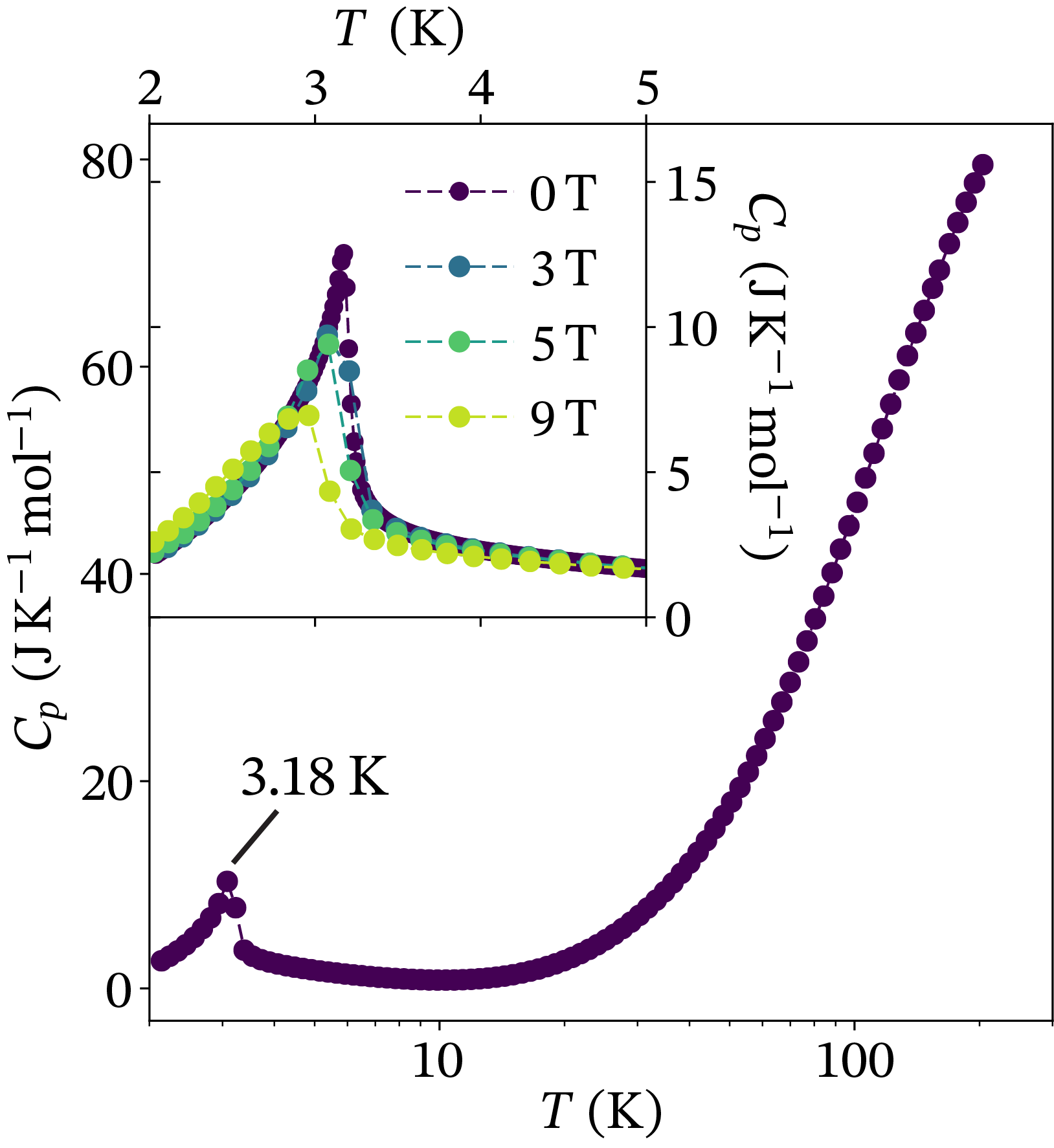}
	\caption{Specific heat $C_p(T)$ of NaCeO$_2$ measured under $\mu_0H = 0, 3, 5, 9$ T fields. The onset of antiferromagnetic order occurs at $T_N = 3.18$ K.}
	\label{fig:figcp}
\end{figure}

Magnetic susceptibility and isothermal magnetization data from NaCeO$_2$ were collected and analyzed between $2 - 300$ K, and data are plotted in Figure \ref{fig:figmagnet}. Data below 200 K and above $T_N$ were fit to a Curie-Weiss behavior, and constants $\mu_{eff} = 0.994 \mu_B = \sqrt{8 C}$, $\Theta_{CW} = -7.69$ K, and $\chi_0 = 0.0022$ emu mol$^{-1}$ were extracted from the fit. A peak in the low-temperature magnetic susceptibility indicative of magnetic ordering was observed at $T_N=3.4$ K with a corresponding inflection in $d(\chi T)/dT$ at 3.3 K, below which ZFC and FC measurements split. In the ordered state, NaCeO$_2$ does not saturate up to fields of $\mu_0H = 14$ T as shown in Figure \ref{fig:figmagnet}b. Instead, it reaches approximately $0.2 \mu_B$, corresponding to only 35\% saturation of the expected moment obtained from magnetic structure data detailed in the next section. 

Low-temperature specific heat data were also collected between 2 \--- 300 K in external magnetic fields of $\mu_0$H $= 0, 3, 5,$and $9$ T. As shown in Figure \ref{fig:figcp}, a sharp transition in the zero field data occurs at $T_N = 3.18$ K corresponding to the antiferromagnetic transition observed in magnetization measurements. This $C_p(T)$ anomaly softens and lowers in temperature as an increasing external magnetic field partially polarizes the Ce moments.  We note here that there is reasonably good agreement between  $d(\chi T)/dT$ obtained from susceptibility measurements and the zero field specific heat \cite{fisher1962relation}, where zero field $C_p(T)$ data reveal $T_N = 3.18$ K close to $d(\chi T)/dT =$ 3.3 K determined via $\mu_0$H $= 50$ Oe susceptibility . 

\subsection{Crystal structure}

\begin{figure}[t]
	\includegraphics[scale=.6]{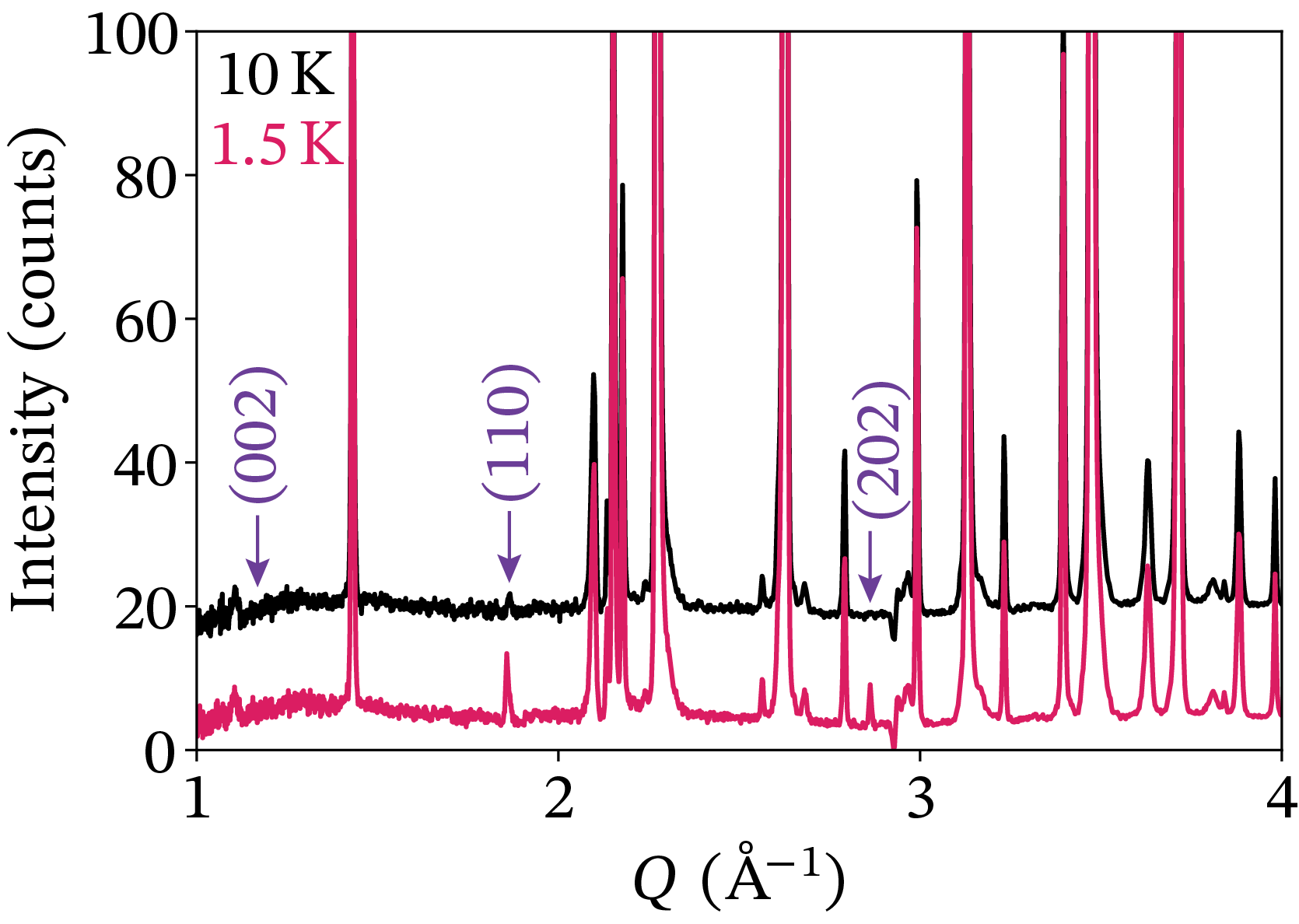}
	\caption{Neutron diffraction data collected on POWGEN in Frame 3 at 10 K (black) and 1.5 K (pink) distinctly shows two magnetic reflections indexed at $(1,1,0)$ and $(2,0,2)$ appearing at 1.5 K. The absence of intensity at the $(0,0,2)$ position at 1.5 K indicates that Ce moments align along the $c$-axis.}
	\label{fig:fig002}
\end{figure}

\begin{figure}[]
	\includegraphics[scale=.475]{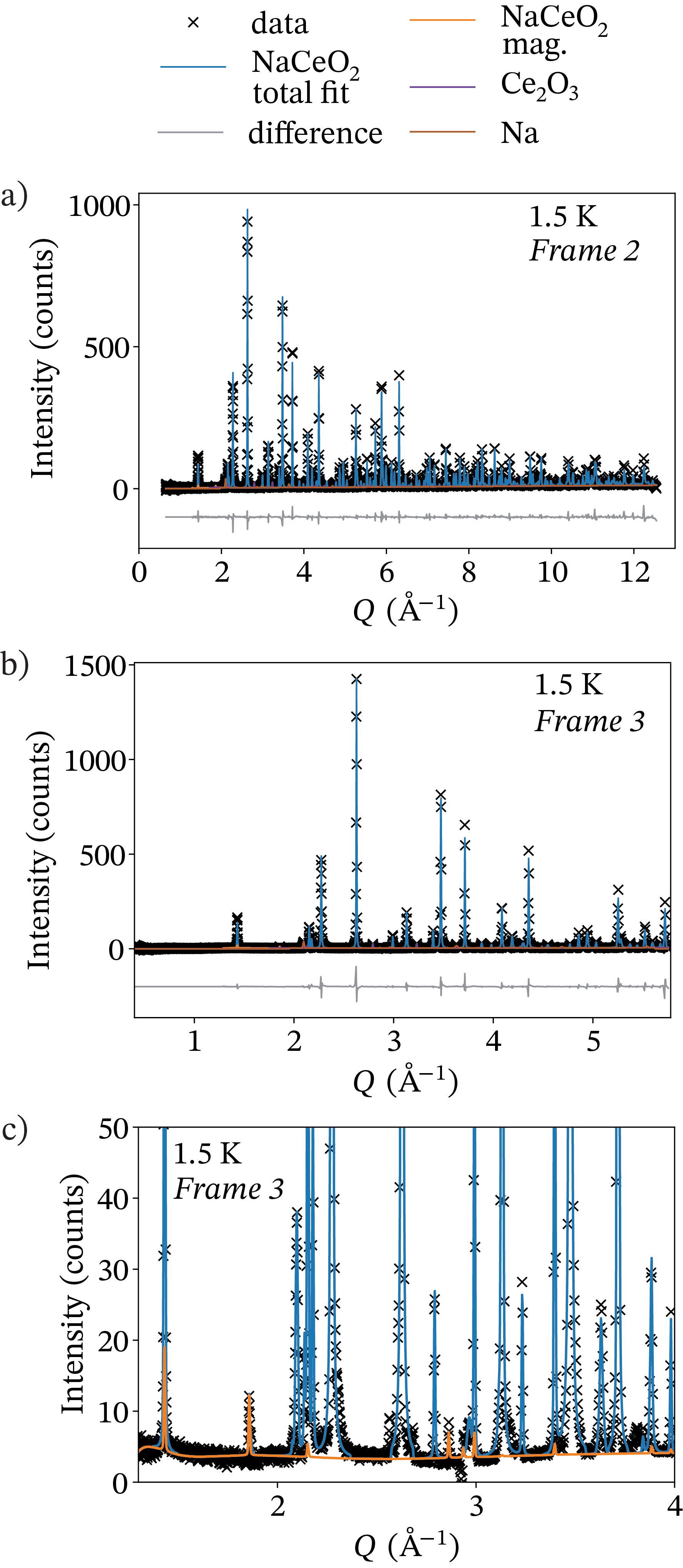}
	\caption{Elastic neutron diffraction of NaCeO$_2$ collected on POWGEN at 1.5 K in a) Frame 2 ($\lambda = 1.5$ \AA) \ and b) Frame 3 ($\lambda = 2.665$ \AA). Data from both frames were co-refined to produce the final fit of the data. NaCeO$_2$ structural parameters are summarized in Table \ref{tab:tabstruct}. Small impurities of 2.2(1)\% of Ce$_2$O$_3$ and 2.1(2)\% of Na by weight were present. c) A closer look at the total fit and data presented in Frame 3. The refined magnetic structure is an $A$-type antiferromagnet with an ordered moment of $0.57(2) \mu_B$.}
	\label{fig:figPND}
\end{figure}

Elastic neutron powder diffraction data collected at 10 K and 1.5 K are shown in Figures \ref{fig:fig002} \--- \ref{fig:figPND}. Results of Rietveld refinements to the structure of NaCeO$_2$ in the second origin choice of $I4_1/amd$ are presented in Table \ref{tab:tabstruct}. We detect roughly 2.2(1)\% of Ce$_2$O$_3$ and 2.1(2)\% of Na by weight as impurity phases in the diffraction pattern, likely as a result of degradation of NaCeO$_2$ from trace water and oxygen exposure during sample transport. The primary NaCeO$_2$ phase shows fully occupied Na, Ce, and O sites in Table \ref{tab:tabstruct}. 

Following the radius ratio rule for $A$Ln$X_2$ ($A$ = alkali; Ln = lanthanide; $X$ = chalcogenide) materials, NaCeO$_2$ crystallizes in the $I4_1/amd$ space group structure \cite{LYOpaper}. This structure contains $D_{2d}$ CeO$_6$ distorted, edge-sharing octahedra in a bipartite tetragonally-elongated diamond lattice. Nearest-neighbor Ce \--- Ce distances span $J_1$ (3.65105 \AA) and next-nearest-neighbor Ce \--- Ce distances are coupled by $J_2$ (4.77860 \AA) in Figure \ref{fig:figstruct}b. Despite their relatively large difference of roughly 1.1 \AA, oxygen mediated superexchange is promoted along $J_2$ with a Ce-O-Ce bond angle at 164$^{\circ}$ relative to $J_1$ where the Ce-O-Ce bond angle is 98$^{\circ}$. In fact, the Ce-O-Ce distance of $J_1$ and $J_2$ are relatively similar at 4.834 \AA \ and 4.827 \AA, respectively. A similar situation has been observed in the related material LiYbO$_2$, where, in the Heisenberg $J_1$\---$J_2$ limit, the bipartite Yb lattice becomes geometrically frustrated when $J_1$ and $J_2$ compete and can form a variety of ferromagnetic, antiferromagnetic, and spiral magnetic phases \cite{LYOpaper}.

We also note that in the $I4_1/amd$ crystallographic structure, the $(1,1,0)$ reflection is naively forbidden. However, weak intensity at the $(1,1,0)$ position appears in the 10 K data in Figure \ref{fig:fig002}, suggesting either a weak violation of the $I4_1/amd$ space group in the nuclear structure of NaCeO$_2$ or that a portion of the NaCeO$_2$ powder had not fully thermalized at 10 K after warming from 1.5 K. For this analysis, this weak violation was ignored, and both nuclear and magnetic structures were analyzed in the ideal $I4_1/amd$ space group. 

\begin{table}[]
	\caption{Rietveld refinement of structural parameters at 1.5\,K from co-refined POWGEN Frame 2 and Frame 3 elastic neutron scattering data. NaCeO$_2$ was modeled in the $I4_1/amd$ space group using origin setting 2. Within error, all ions refine to full occupation and no site mixing is observed.}
	\begin{tabular}{cc|ccccc}
		\hline
		\multicolumn{2}{c|}{$T$}       & \multicolumn{5}{c}{1.5 K}     \\ \hline
		\multicolumn{2}{c|}{$a=b$}     & \multicolumn{5}{c}{4.77860(3) \AA}  \\
		\multicolumn{2}{c|}{$c$}       & \multicolumn{5}{c}{11.04277(11) \AA}  \\ \hline
		Atom           & Wyckoff          & x    & y    & z            & $B_{iso}$ (\AA$^{2}$)      & Occupancy  \\ \hline
		Ce             & 4a            & 0    & 0    & 0            & 0.057(20)   & 0.992(5) \\
		Na             & 4b            & 0    & 0    & 0.5          & 0.564(20)   & 1.000(4) \\
		O              & 8e            & 0    & 0    & 0.21921(9)   & 0.284(11)   & 1.000(2) \\ \hline	
	\end{tabular}
	\label{tab:tabstruct}
	
\end{table}

\subsection{Magnetic structure}

Neutron diffraction data collected on NaCeO$_2$ further show that new magnetic reflections develop below $T_N =$ 3.18 K as shown in Figure \ref{fig:fig002}. Examining the 1.5 K data reveals two distinct magnetic reflections indexed to \textbf{Q}=$(1,1,0)$ and \textbf{Q}=$(2,0,2)$. The magnetic reflections at 1.5 K were therefore indexed to a commensurate ordering wave vector ${\bf{k}} = (0,0,0)$. No discernible intensity arises at the \textbf{Q}=$(0,0,2)$ position, indicating the spins lay parallel to the $c$-axis in the ordered phase. The antiferromagnetic structure is generated by the $\Gamma_7$ irreducible representation in $I4_1/amd$ and is shown in Figure \ref{fig:figstruct}b. Moments within the $ab$-plane coalign and moments along the $c$-axis antialign in an A-type pattern of antiferromagnetic order in the $I4_1'/a'm'd$ magnetic space group. Similar magnetic reflection conditions and structures have been observed in other antiferromagnetic materials sharing the $I4_1/amd$ or related $I4_1/a$ space group (e.g. YbVO$_4$, KRuO$_4$ \cite{yvo, kro}). Rietveld refinement of the magnetic structure using this model reveals an ordered Ce$^{3+}$ moment of 0.57(2) $\mu_B$. This indicates an average g-factor $g_{avg}$=1.14, a value consistent with independent susceptibility measurements of $g_{avg}$=1.15 in Figure \ref{fig:figmagnet}c.

\section{V. Discussion}

NaCeO$_2$ provides as an important example of a commensurate magnetically ordered state in the phase diagram of $J_{eff} = 1/2$ moments decorating the tetragonally-elongated diamond lattice of $ALnX_2$ compounds. This commensurate phase differs from a recent report studying the magnetic ground state in LiYbO$_2$ \cite{LYOpaper}, where it was reported that LiYbO$_2$ forms an incommensurate spiral magnetic phase in zero field below $T_{AF}\approx1$ K with an ordering wave vector ${\bf{k}} = (0.384, \pm 0.384, 0)$. 

The magnetic order in LiYbO$_2$ was previously analyzed by adapting the frustrated diamond lattice Heisenberg $J_1$\---$J_2$ model \cite{bergman-balents, lee-balents, buessen-trebst, bernier2008quantum, gchen} in the extreme limit of tetragonal distortion.  Here the ordering wave vector is uniquely determined by the ratio of $J_2/|J_1|$. The same tetragonal Heisenberg model also predicts a commensurate, N\'eel phase in the limit where $J_1 > 4 J_2$ and $J_1 > 0$. This commensurate magnetic state coincides with the structure determined for NaCeO$_2$ and suggests that the antiferromagnetic $J_1$ term dominates over $J_2$ in NaCeO$_2$. The ratio of these two exchange energies can therefore seemingly be tuned via relatively small lattice perturbations and by chemically alloying across the tetragonal variants of the $ALnX_2$ series.  

The precise mapping of exchange interactions between the two systems, LiYbO$_2$ and NaCeO$_2$, can be modified by anisotropies and other interactions not captured in the minimal Heisenberg $J_1-J_2$ Hamiltonian presented in our earlier work \cite{LYOpaper}.  For instance, an XXZ anisotropy can alter the relative phase boundaries since the effect of anisotropy is to renormalize the effective exchange interaction strengths and their ratios.  Nevertheless, structural changes between NaCeO$_2$ and LiYbO$_2$, such as relative changes in the $Ln$-O-$Ln$ $J_1$ bond angles and ratios of $Ln$-O bond lengths for $J_1$ and $J_2$ pathways, naively should promote a larger $J_1/J_2$ ratio in NaCeO$_2$. This trend and the change in the magnetic ground state are qualitatively consistent with the expectations of the idealized $J_1-J_2$ model.

Knowing that spiral and commensurate phases occur in LiYbO$_2$ and NaCeO$_2$, respectively, the entire family of $I4_1/amd$ $ALnX_2$ materials potentially represent a unique opportunity to study the frustrated Heisenberg model in the elongated $J_1$\---$J_2$ limit as a function of $A$ and $Ln$-ion tunability. Reports have shown that Li$Ln$O$_2$ ($Ln =$ Sc, Lu, Er) \cite{LiLnO2cmpds} and Na$Ln$O$_2$ ($Ln =$ Nd, Sm, Eu,  Gd) \cite{NaLnO2cmpds} also crystallize in this space group. Some of these materials display sharp antiferromagnetic transitions in specific heat measurements like NaCeO$_2$, while others show broad anomalies like LiYbO$_2$ \cite{LiLnO2cmpds, LYOpaper} or even ferromagnetic transitions like NaNdO$_2$ \cite{NaLnO2cmpds}. Future work understanding the impact of varying the lanthanide ion character and single-ion/exchange anisotropies is an appealing next step for further refining the Heisenberg $J_1$\---$J_2$ model for these compounds.

\section{VI. Conclusions}

The magnetic ground state and crystalline-electric field Hamiltonian of Ce-ions in the tetragonally elongated diamond lattice of NaCeO$_2$ was determined. This material crystallizes in the $I4_1/amd$ structure type of the $ALnX_2$ family of compounds, and heat capacity and magnetization measurements show that NaCeO$_2$ develops long-range magnetic order below $T_N = 3.18$ K. New magnetic reflections appear in neutron powder diffraction data and reveal A-type antiferromagnetic order with an ordered moment of 0.57(2) $\mu_B$ per Ce ion. The crystalline-electric field scheme of the ground state $J = 5/2$ multiplet was determined, and the ground state wave function was determined to be of mixed $m_z=3/2$ and $m_z=5/2$ character. When mapped onto a Heisenberg $J_1$\---$J_2$ model, the commensurate antiferromagnetic order observed in this system implies an enhancement in the ratio of effective exchange parameters $J_1/J2$ relative to the spiral state formed in LiYbO$_2$. 

\section{Acknowledgments}
\begin{acknowledgments}
	S.D.W. and M.B. sincerely thank Leon Balents and Chunxiao Liu for helpful discussions.  This work was supported by the US Department of Energy, Office of Basic Energy Sciences, Division of Materials Sciences and Engineering under award DE-SC0017752 (S.D.W. and M.B.).  Research reported here also made use of shared facilities of the UCSB MRSEC (NSF DMR-1720256). A portion of this research used resources at the Spallation Neutron Source, a DOE Office of Science User Facility operated by Oak Ridge National Laboratory.
\end{acknowledgments}


\bibliography{NCO_paper1_bib_v2}

\end{document}